\documentclass{article}
\usepackage{amssymb}
\usepackage{amsmath}
\usepackage{amsthm}
\usepackage{mathtools}
\usepackage{todonotes}
\usepackage{hyperref}
\usepackage{enumerate}
\usepackage{caption}
\usepackage{subcaption,graphicx}
\usepackage{geometry}
\usepackage{tikz}
\usepackage{pgfplots}
\usepgfplotslibrary{external}
\usepackage[ruled]{algorithm2e}
\usepackage[page]{appendix}

\newtheorem{theorem}{Theorem}

\newtheorem{remark}[theorem]{Remark}

\theoremstyle{definition}

\makeatletter
\renewcommand{\todo}[2][]{\tikzexternaldisable\@todo[#1]{#2}\tikzexternalenable}
\makeatother

\usepackage{xcolor}

\title{Denoising Gradient Descent in Variational Quantum Algorithms}
\author{\hspace*{-2em}Lars Simon \\ \hspace*{-1em}Bundesdruckerei GmbH \\ {\tt\hspace*{-1em} lars.simon@bdr.de} \and \hspace*{5.em}Holger Eble \\ \hspace*{5.2em}Bundesdruckerei GmbH \\ {\tt \hspace*{4.9em}holger.eble@bdr.de}  \and \hspace*{-4.25em}Hagen-Henrik Kowalski \\ \hspace*{-4.1em}Bundesdruckerei GmbH \\ {\tt \hspace*{-2em}hagen-henrik.kowalski@bdr.de}  \and \hspace*{1.6em}Manuel Radons \\\hspace*{1.5em} Bundesdruckerei GmbH \\ {\tt\hspace*{1.5em} manuel.radons@bdr.de}}
\date{March 2024}

\begin{document}
	
	\maketitle
	
	\begin{abstract}
		In this article we introduce an algorithm for mitigating the adverse effects of noise on gradient descent in variational quantum algorithms. This is accomplished by computing a {\emph{regularized}} local classical approximation to the objective function at every gradient descent step. The computational overhead of our algorithm is entirely classical, i.e., the number of circuit evaluations is exactly the same as when carrying out gradient descent using the parameter-shift rules. We empirically demonstrate the advantages offered by our algorithm on randomized parametrized quantum circuits.
	\end{abstract}

	\section{Introduction}
    In variational quantum algorithms (VQAs) a (typically gradient-based) classical optimizer is used to train a parametrized quantum circuit. While there is evidence that the presence of noise can be helpful for avoiding saddle points in VQAs \cite{VQA_noise_saddle_points}, noise is generally detrimental to their performance \cite{noise_VQA}, \cite{noise_induced_barren_plateaus}. In line with this, several techniques for mitigating the effect of noise in quantum algorithms (and VQAs in particular) have been proposed, see, e.g., \cite{variational_quantum_simulator_error_minimization}, 
    \cite{error_mitigation_short_circuits}, 
    \cite{Czarnik2021errormitigation}, 
    \cite{readout-error_mitigation_vdBerg},
    \cite{depolarizing_noise_mitigation_Urbanek},
    \cite{VQA_error_mitigation}.

    In this article we introduce an algorithm for mitigating the adverse effects of noise on gradient descent in VQAs. As is the case for the algorithm introduced in \cite{VQA_error_mitigation}, the error mitigating techniques are applied in real-time, i.e., during execution of gradient descent. The idea is, roughly speaking, to compute an approximation to the objective function at every gradient descent step. Computation of such approximations is facilitated by the fact that the set of possible objective functions naturally embeds into a certain reproducing kernel Hilbert space, whose structure can be exploited. Non-surprisingly, the quality of the approximation will in general not be good on the entire parameter space – otherwise we could forego the need for a quantum device and simply work with the classical approximation instead. However, at every gradient descent step, we are guaranteed to obtain a good approximation {\emph{locally}} around the current point in parameter space, see Remark \ref{remark:no_noise_means_exact_equality} below. The benefit of computing a local approximation is that samples from past iterations can be taken into account in order to make the approximation more robust to noise. Moreover, computing an approximation allows us to use regularization techniques from classical machine learning.
    
    Our method is agnostic to the type of noise that evaluation of the objective function is subjected to; as a result, our algorithm can be seen as a general purpose method with applications in a variety of settings. However, this agnosticism comes with some disadvantages, the most obvious one being that our algorithm might fail to mitigate certain types of noise. Moreover, for specific types of noise, our algorithm is likely to be outperformed by specialized methods. The advantages and drawbacks of our algorithm are discussed in more detail in Section \ref{sec:discussion}.

    Gradients in VQAs are usually calculated using the so-called parameter-shift rules \cite{Mitarai_2018}, \cite{Schuld_2019}, \cite{Mari_2021}, \cite{Wierichs2022generalparameter} (although the computational cost of the latter scales very unfavourably with the number of trainable parameters, see \cite{VQA_backpropagation_scaling}, \cite{abbas_backpropagation_state_tomography}). The computational overhead of our algorithm is entirely classical, i.e., the number of circuit evaluations is exactly the same as when carrying out gradient descent using the parameter-shift rules. 

	This article is organized as follows. In Section \ref{sec:algorithm} we introduce our algorithm, which includes a description in pseudocode, see Algorithm \ref{algorithm:denoised_gradient_descent}. In Section \ref{sec:experiments} we analyse our algorithm experimentally on a large number of randomized parametrized quantum circuits, considering both measurement shot noise and (simulated) quantum hardware noise by using some of the fake backends (designed to mimic the behavior of IBM Quantum systems) provided by the Qiskit framework. In Section \ref{sec:discussion} we discuss the advantages and the drawbacks of our algorithm. Finally, in Section \ref{sec:conclusion}, we make our concluding remarks and point out some possible directions for future research.

    \subsection{Acknowledgement}
    This article was written as part of the Qu-Gov project, which was commissioned by the German Federal Ministry of Finance. The authors want to extend their gratitude to Kim Nguyen, Manfred Paeschke, Oliver Muth, Andreas Wilke, and Yvonne Ripke for their continuous encouragement and support.
 
	\section{The Algorithm}
	\label{sec:algorithm}
    In Section \ref{subsec:algorithm_setting} we describe the setting for our algorithm and the problem at hand, and in Section \ref{subsec:algorithm_denoised} we describe the algorithm we developed to tackle the problem.
    
    \subsection{The Setting}
    \label{subsec:algorithm_setting}
	We let $n$ be a positive integer and, for $\theta\in\mathbb{R}^m$, consider the unitary
	\begin{align*}
		U(\theta)
		=C_{m+1}R_m (\theta_m) C_m\cdots R_2 (\theta_2) C_2 R_1 (\theta_1) C_1 
		\in\mathbb{C}^{2^n\times 2^n},
	\end{align*}
	where $m$ is a non-negative integer, $C_1 ,\dots , C_{m+1}$ are unitaries given by $n$-qubit quantum circuits and, for all $j\in\{1,\dots ,m\}$, the unitary $R_j (\theta_j)$ is a rotation of the form 
	\begin{align*}
		R_j (\theta_j) = 
		\exp \left(-i\frac{\theta_j}{2}G_j\right)
		\in \mathbb{C}^{2^n\times 2^n}
	\end{align*}
	for some Hermitian $G_j\in \mathbb{C}^{2^n\times 2^n}$ whose set of Eigenvalues is $\{-1,1\}$. For example, each $G_j$ could be a tensor product $P_1\otimes\cdots\otimes P_n\neq I^{\otimes n}$ of Pauli matrices $P_1,\dots ,P_n\in\{I,X,Y,Z\}$. Moreover, we consider an observable given by a Hermitian matrix $\mathcal{M}\in\mathbb{C}^{2^n\times 2^n}$. Letting $|\psi (\theta)\rangle :=U(\theta)|0\rangle^{\otimes n}$, we consider the expected value
	\begin{align*}
		f(\theta):=\langle\psi (\theta)|\mathcal{M}|\psi (\theta)\rangle
		,
	\end{align*}
	which defines a function $f\colon\mathbb{R}^m\to\mathbb{R}$. The aim is now to minimize $f$ using gradient descent. Gradients of $f$ are typically calculated using the so-called parameter-shift rules \cite{Mitarai_2018}, \cite{Schuld_2019}, \cite{Mari_2021}, which reduce computation of the gradient to $2m$ evaluations of $f$. However, evaluation of $f$ is subject to both measurement shot noise and hardware quantum noise, since it involves the execution of quantum circuits on a quantum device. As a result, gradients obtained in this way are themselves noisy, which is detrimental to performance of the gradient descent algorithm.
	
	\subsection{Denoised Gradient Descent Algorithm}
    \label{subsec:algorithm_denoised}

	As shown in \cite{Schuld_2021}, we can write
	\begin{align*}
		f(\theta) =\sum_{\omega\in\{-1,0,1\}^m} c_\omega e^{i\omega^t \theta}
		\text{ for all }\theta\in\mathbb{R}^m,
	\end{align*}
	where $c_\omega\in\mathbb{C}$ and $c_\omega = \overline{c_{-\omega}}$ for all $\omega\in\{-1,0,1\}^m$. We denote the set of all functions of this form as $H$ and give $H$ the structure of a reproducing kernel Hilbert space with reproducing kernel $K$ by defining
	\begin{align*}
		\langle g_1 ,g_2\rangle_H= \int_{[-\pi ,\pi]^m} g_1 (z) g_2 (z) \mathrm{d}z, \ g_1,g_2\in H,
	\end{align*}
	and
	\begin{align*}
		K(x,z) = \frac{1}{(2\pi )^m} \prod_{j=1}^m (1+2\cos(x_j - z_j )), \ x,z\in\mathbb{R}^m.
	\end{align*}
    For the sake of numerical stability we will consider $\tilde{K}$ instead of $K$ in our algorithms, where the former is given by
    \begin{align*}
			\tilde{K}(x,z) = \prod_{j=1}^m \frac{1+2\cos(x_j - z_j )}{3}, \ x,z\in\mathbb{R}^m.
		\end{align*}
	A more rigorous treatment of the above can be found in \cite{simon2023interpolating}.

    We will now describe the algorithm. As is the case for the gradient descent algorithm, we start out with an initial point $\theta_0\in\mathbb{R}^m$, a learning rate $\alpha >0$, and a number of steps $T\in\mathbb{Z}_{\geq 1}$. In addition, we choose a regularization hyperparameter $\lambda >0$ and, optionally, an integer $\ell \geq 1$ for bounding the size of the linear systems of equations we will have to solve. In practice we may, of course, change $\alpha$, $\lambda$, and $\ell$ from iteration to iteration and combine our algorithm with, for example, normalized gradient descent \cite{Hazan2015BeyondCS}, Nesterov’s Accelerated Gradient method \cite{Nesterov1983AMF}, or the ADAM optimizer \cite{KingBa15}, but we will not discuss these options here for the sake of simplicity (for a treatment of these in the context of variational quantum algorithms, see \cite{Suzuki_2021}).

    Now assume that $t\in\mathbb{Z}$, $1\leq t\leq T$ and that we have already carried out $t-1$ steps, which yielded points $\theta_0,\dots , \theta_{t-1}\in\mathbb{R}^m$ in parameter space, and, for $s\in\{0,1,\dots, t-2\}$, (noisy) samples
    \begin{align*}
        g^{(s)}_{j, \nu} \approx f\left(\theta_s +\nu\frac{\pi}{2}\mathbf{e}_j\right), \ j\in\{1,\dots ,m\},\nu\in\{-1,1\},
    \end{align*}
    obtained from evaluating $f\left(\theta_s +\nu\frac{\pi}{2}\mathbf{e}_j\right)$ on a quantum device (note that $\{0,1,\dots, t-2\}$ is the empty set when $t=1$, i.e., in this case we merely start out with $\theta_0$). Here, $\mathbf{e}_j\in\mathbb{R}^m$ denotes the $j^{\text{th}}$ canonical basis vector. We now start the $t^{\text{th}}$ step of the algorithm by first obtaining (noisy) samples $g^{(t-1)}_{j, \nu} \approx f\left(\theta_{t-1} +\nu\frac{\pi}{2}\mathbf{e}_j\right)$, $j\in\{1,\dots ,m\}$, $\nu\in\{-1,1\}$, through evaluation on a quantum device. Instead of using these samples to get a noisy estimate for the gradient $\nabla f(\theta_{t-1})$ by way of the parameter-shift rules, we will instead compute a classical approximation $\tilde{f}_t$ of $f$ and then obtain $\theta_t$ by carrying out a gradient descent step with respect to $\tilde{f}_t$. Non-surprisingly, the quality of the approximation will in general not be good on the entire parameter space $\mathbb{R}^m$ -- otherwise we could forego the need for a quantum device and simply work with the classical approximation instead. However, we are guaranteed to obtain a good estimate for the gradient of $f$ at the point $\theta_{t-1}$, see Remark \ref{remark:no_noise_means_exact_equality}.

    In order to construct the approximation $\tilde{f}_t$, we will make use of the (noisy) samples of $f$ we have obtained so far. For ease of notation, we denote the collection of points $\left(\theta_s +\nu\frac{\pi}{2}\mathbf{e}_j\right)_{s, j, \nu}$ as $p^{(t)}_1 ,\dots , p^{(t)}_{D_t}$ and the corresponding noisy samples $\left(g^{(s)}_{j, \nu}\right)_{s, j, \nu}$ as $v^{(t)}_1 ,\dots , v^{(t)}_{D_t}$ (the order in which we list the points does not have an impact on the function $\tilde{f}_t$ we are constructing). Here, $j\in\{1,\dots ,m\}$, $\nu\in\{-1,1\}$, and $s$ ranges from $0$ to $t-1$ if the optional input $\ell$ was not provided. If the optional input $\ell$ was provided, we will discard some of the older samples in order to ensure that it does not become infeasible to compute the approximation $\tilde{f}_t$. In this case, $s$ ranges from $\max\{0, t-\ell\}$ to $t-1$. We now consider the linear system of equations
    \begin{align*}
		\left(\left(\tilde{K}(p^{(t)}_k ,p^{(t)}_l )\right)_{1\leq k,l\leq D_t}+\lambda I_{D_t\times D_t}\right)\cdot\eta =
		\begin{pmatrix}
			v^{(t)}_1 \\
			\vdots\\
			v^{(t)}_{D_t}
		\end{pmatrix}
		, \text{ where } \eta\in\mathbb{R}^{D_t} ,
	\end{align*}
    where $I_{D_t\times D_t}$ denotes the ${D_t\times D_t}$ identity matrix. Since $\tilde{K}$ only differs from the symmetric positive semidefinite kernel $K$ by multiplication with a positive constant, the matrix appearing in this linear system of equation is strictly positive definite (recall that $\lambda >0$) and hence invertible. This implies that there exists a uniquely determined solution $\eta^{(t)}\in\mathbb{R}^{D_t}$. We now define $\tilde{f}_t\colon\mathbb{R}^m\to\mathbb{R}$ by
    \begin{align*}
        \tilde{f}_t (\theta) =\sum_{k=1}^{D_t} \eta^{(t)}_k \tilde{K}(p^{(t)}_k, \theta)
		\text{ for all }\theta\in\mathbb{R}^m .
    \end{align*}
    Note that $\tilde{f}_t$ can be evaluated on a classical device. Since $\tilde{f}_t\in H$, the parameter-shift rules apply (see Lemma 10 in \cite{simon2023interpolating}), and hence we get, for all $j\in\{1,\dots ,m\}$:
    \begin{align*}
        \partial_j \tilde{f}_t (\theta_{t-1}) = \frac{\tilde{f}_t\left(\theta_{t-1} +\frac{\pi}{2}\mathbf{e}_j\right)-\tilde{f}_t\left(\theta_{t-1} -\frac{\pi}{2}\mathbf{e}_j\right)}{2}
        ,
    \end{align*}
    which allows us to compute $\nabla \tilde{f}_t (\theta_{t-1})$. Alternatively, $\nabla \tilde{f}_t (\theta_{t-1})$ can be computed using the explicit expression for $\tilde{K}$ given above. We now obtain $\theta_t$ as
    \begin{align*}
        \theta_t = \theta_{t-1}-\alpha\cdot \nabla \tilde{f}_t (\theta_{t-1})
        .
    \end{align*}
    The use of the regularization parameter $\lambda$ also has an impact on the length of the gradient vector $\nabla \tilde{f}_t (\theta_{t-1})$ and we are, arguably, more interested in denoising the {\textit{direction}} of the estimate for the gradient vector of $f$. Because of this, we may optionally want to force the step length to be the same as it would be when using the noisy estimate for the gradient of $f$. One way of doing this would be to use the alternative descent step $\theta_t = \theta_{t-1}-\alpha_t\cdot \nabla \tilde{f}_t (\theta_{t-1})$, where
    \begin{align*}
        \alpha_t = \frac
        {\left\Vert \frac{1}{2}
        \begin{pmatrix}
            g^{(t-1)}_{1, 1}-g^{(t-1)}_{1, -1}\\
            \vdots\\
            g^{(t-1)}_{m, 1}-g^{(t-1)}_{m, -1}
        \end{pmatrix}
        \right\Vert+\epsilon}
        {\left\Vert\nabla \tilde{f}_t (\theta_{t-1})\right\Vert +\epsilon}\cdot \alpha
        .
    \end{align*}
    Here, $\Vert\cdot\Vert$ denotes the Euclidean norm and we introduced a small constant $\epsilon >0$ for the sake of numerical stability and in order to avoid division by $0$. This concludes our description of the algorithm. Note that the computational overhead of our algorithm is entirely classical; our algorithm needs exactly as many circuit evaluations as gradient descent.
    
    The intuition behind the algorithm is explained in Remark \ref{remark:no_noise_means_exact_equality}. A compact description of our algorithm in pseudocode can be found in Algorithm \ref{algorithm:denoised_gradient_descent}.

    	\begin{algorithm}
		\SetKwInOut{Input}{Input}
		\SetKwInOut{Output}{Output}
        \SetKwInOut{Hyperparameters}{Hyperparameters}

        \Hyperparameters{learning rate $\alpha >0$, regularization $\lambda >0$, {\textit{optional:}} bound $\ell\in\mathbb{Z}_{\geq 1}$ on number of iterations to consider when constructing approximation, {\textit{optional:}} $\epsilon >0$ for numerical stability when rescaling gradient}
  
		\Input{initial point $\theta_0\in\mathbb{R}^m$, number of steps $T\in\mathbb{Z}_{\geq 1}$, function $f\colon\mathbb{R}^m\to\mathbb{R}$, $f(\theta)= \langle0^n|U^\dag (\theta)\mathcal{M}U(\theta)|0^n\rangle$ (see Section \ref{subsec:algorithm_setting})}
		\Output{point $\theta_T\in\mathbb{R}^m$ in parameter space}

        \nl \If{optional hyperparameter $\ell$ not provided}
        {
            \nl Set $\ell :=T$
        }
        
        \nl \For{$t=1,\dots ,T$}
		{
			\nl \For{$j=1,\dots ,m$}
    		{
    			\nl Obtain (noisy) samples
                \begin{align*}
                    g^{(t-1)}_{j, 1} \approx f\left(\theta_{t-1} +\frac{\pi}{2}\mathbf{e}_j\right),
                    &&
                    g^{(t-1)}_{j, -1} \approx f\left(\theta_{t-1} -\frac{\pi}{2}\mathbf{e}_j\right),
                \end{align*}
                by evaluating $f$ at the respective points using a quantum device
    		}
            \nl Assemble the collections
            \begin{align*}
                \left(\theta_s +\nu\frac{\pi}{2}\mathbf{e}_j\right)_{s, j, \nu},
                &&
                \left(g^{(s)}_{j, \nu}\right)_{s, j, \nu}
            \end{align*}
            into tuples $\left({p^{(t)}_1 ,\dots , p^{(t)}_{D_t}}\right)$ and $\left({v^{(t)}_1 ,\dots , v^{(t)}_{D_t}}\right)$ respectively, in arbitrary (but matching) order, where $j\in\{1,\dots ,m\}$, $\nu\in\{-1,1\}$, and $s\in\{\max\{0, t-\ell\}, \dots, t-1\}$

            \nl Find the uniquely determined $\eta^{(t)}\in\mathbb{R}^{D_t}$ solving the linear system of equations
            \begin{align*}
        		\left(\left(\tilde{K}(p^{(t)}_k ,p^{(t)}_l )\right)_{1\leq k,l\leq D_t}+\lambda I_{D_t\times D_t}\right)\cdot\eta =
        		\begin{pmatrix}
        			v^{(t)}_1 \\
        			\vdots\\
        			v^{(t)}_{D_t}
        		\end{pmatrix}
        		, \text{ where } \eta\in\mathbb{R}^{D_t}
        	\end{align*}

            \nl Set $\tilde{f}_t (\theta) =\sum_{k=1}^{D_t} \eta^{(t)}_k \tilde{K}(p^{(t)}_k, \theta)$

            \nl Compute the gradient
            \begin{align*}
                \nabla \tilde{f}_t (\theta_{t-1}) = 
                \left(
                {\frac{\tilde{f}_t\left(\theta_{t-1} +\frac{\pi}{2}\mathbf{e}_j\right)-\tilde{f}_t\left(\theta_{t-1} -\frac{\pi}{2}\mathbf{e}_j\right)}{2}}
                \right)_{j=1,\dots ,m}
            \end{align*}
            (alternatively, $\nabla \tilde{f}_t (\theta_{t-1})$ can be computed using the explicit expression for $\tilde{K}$)

            \nl \If{optional hyperparameter $\epsilon$ not provided}
                {\nl Set $\alpha_t :=\alpha$ (in this case we do not rescale the gradient)}
                \nl \Else
                {\nl Set $\alpha_t :=\frac
                    {\left\Vert \frac{1}{2}
                    \begin{pmatrix}
                        g^{(t-1)}_{1, 1}-g^{(t-1)}_{1, -1}\\
                        \vdots\\
                        g^{(t-1)}_{m, 1}-g^{(t-1)}_{m, -1}
                    \end{pmatrix}
                    \right\Vert+\epsilon}
                    {\left\Vert\nabla \tilde{f}_t (\theta_{t-1})\right\Vert +\epsilon}\cdot \alpha$}

            \nl Set $\theta_t := \theta_{t-1}-\alpha_t\cdot \nabla \tilde{f}_t (\theta_{t-1})$
		}

        \nl \Return $\theta_T$

		\caption{Denoised Gradient Descent}
		\label{algorithm:denoised_gradient_descent}
	\end{algorithm}

    \begin{remark}\label{remark:no_noise_means_exact_equality}
        In order to understand the significance the function $\tilde{f}_t$, it is instructive to consider the {\textit{hypothetical}} scenario where $\lambda$ is replaced by $0$ and where all samples we obtained are completely noiseless, i.e., coincide with the exact values of $f$ at the respective points. In this case, the above linear system of equations reduces to
        \begin{align*}
    		\left(\tilde{K}(p^{(t)}_k ,p^{(t)}_l )\right)_{1\leq k,l\leq D_t}\cdot\eta =
    		\begin{pmatrix}
    			f(p^{(t)}_1) \\
    			\vdots\\
    			f(p^{(t)}_{D_t})
    		\end{pmatrix}
    		, \text{ where } \eta\in\mathbb{R}^{D_t}.
	    \end{align*}
        As was shown in the appendix of \cite{simon2023interpolating} (see also Implementation Remark 6 in this reference), the linear system of equations still has a solution $\eta^{(t)}\in\mathbb{R}^{D_t}$ in this setting, and the function $\tilde{f}_t$, defined as above, agrees with $f$ on $\{p^{(t)}_1 ,\dots , p^{(t)}_{D_t}\}$. In particular, since the parameter-shift rules apply to both $f$ and $\tilde{f}_t$ (since both are $\in H$), we have for all $j\in\{1,\dots ,m\}$:
        \begin{align*}
            \partial_j f (\theta_{t-1})
            = \frac{f\left(\theta_{t-1} +\frac{\pi}{2}\mathbf{e}_j\right)-f\left(\theta_{t-1} -\frac{\pi}{2}\mathbf{e}_j\right)}{2}
            = \frac{\tilde{f}_t\left(\theta_{t-1} +\frac{\pi}{2}\mathbf{e}_j\right)-\tilde{f}_t\left(\theta_{t-1} -\frac{\pi}{2}\mathbf{e}_j\right)}{2}
            = \partial_j \tilde{f}_t (\theta_{t-1})
            .
        \end{align*}
        It follows that $\nabla f (\theta_{t-1}) = \nabla \tilde{f}_t (\theta_{t-1})$. So, in the noiseless case without regularization hyperparameter, we exactly recover the gradient of $f$ at $\theta_{t-1}$. 

        In the noisy case, we need to introduce a small $\lambda >0$ in order to ensure that the linear system of equations still has a solution. In this case, the function $\tilde{f}_t$ corresponds to a solution to the optimization problem underlying kernel ridge regression \cite{kernel_ridge_regression} with kernel $\tilde{K}$, regularization hyperparameter $\lambda$, and data $\left((p_k^{(t)}, v_k^{(t)})\right)_{k=1,\dots ,D_t}$. One can hope that, in the spirit of Tikhonov regularization \cite{tikhonov_regularization}, the regularization hyperparameter $\lambda$ helps mitigate the effect that noise (stemming from the evaluation of $f$ on a quantum device) has on the solution to the above linear system of equations. Moreover, this approach is very natural, since the true function $f$ is known to be contained in $H$, the reproducing kernel Hilbert space associated to the kernel $K$, which only differs from $\tilde{K}$ by multiplication with a positive constant. However, the main benefit of our method stems from the fact that (when $\ell \geq 2$ or the optional input $\ell$ was not provided), samples from past iterations are used to improve the quality of the classical approximation $\tilde{f}_t$. While one would not expect this to be beneficial in the noiseless case (we recover the exact gradient using the samples from the current iteration, see above), it is plausible that, in the noisy case, the increased number of samples involved in the approximation process will reduce the effect of noise on the gradient estimate provided by our approximation. We give numerical evidence for this in Section \ref{sec:experiments}; a thorough theoretical analysis is left for future work.
    \end{remark}

    \section{Experiments}\label{sec:experiments}
    In this section we describe our experiments with Algorithm \ref{algorithm:denoised_gradient_descent}. In Section \ref{subsec:alignment_exact_grad} we will randomly sample parametrized quantum circuits and points in parameter space and compare the quality of the noisy gradient to that of the denoised gradient computed by Algorithm \ref{algorithm:denoised_gradient_descent}. Here, {\emph{quality}} is measured in terms of cosine similarity to the exact gradient vector, computed via statevector simulation. In Section \ref{subsec:descent_objective} we randomly sample a parametrized quantum circuit and an initial point in parameter space and compare the descent of the objective function between many executions of noisy gradient descent and Algorithm \ref{algorithm:denoised_gradient_descent} respectively.  

\begin{figure}[htp]
	\centering
    {\includegraphics[width=6.7in]{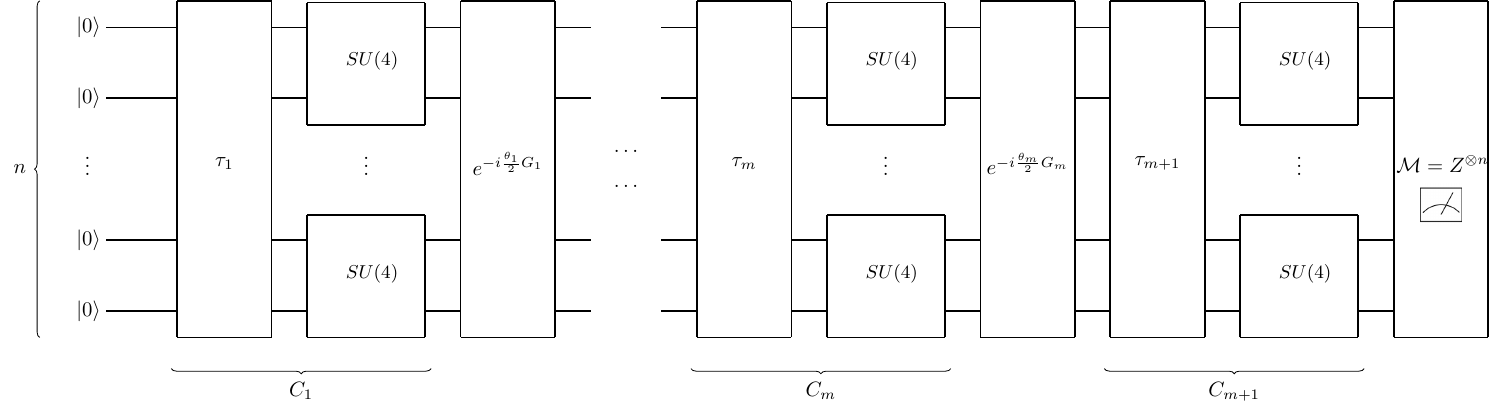}}
	\caption{This figure shows the random circuits used in the experiments in Section \ref{sec:experiments}. Here, $C_1 , \dots , C_{m+1}$ are randomly sampled just like the individual layers in the quantum volume test \cite{quantum_volume}. Moreover, $G_1 ,\dots , G_m$ are randomly sampled non-identity Pauli strings and the measurement observable is always $\mathcal{M}=Z^{\otimes n}$. For a more rigorous description of these circuits, see the beginning of Section \ref{sec:experiments}.}
	\label{fig:circuit}
    \end{figure}

    In all our experiments, points in parameter space were randomly sampled from the uniform distribution on $[0,2\pi )^m$ and we worked with the fixed observable $\mathcal{M} = Z^{\otimes n}$ and initial state $|0\rangle^{\otimes n}$. Informally speaking, we created random circuits by sandwiching random parametrized n-qubit Pauli rotations between random unitaries. More precisely, with the notation from Section \ref{subsec:algorithm_setting}, $G_1, \dots , G_m$ were randomly sampled (independently) from the uniform distribution on $\{I,X,Y,Z\}^{\otimes n}\setminus\{I^{\otimes n}\}$. In order to keep the circuit depth manageable, the unitaries $C_1 ,\dots , C_{m+1}$ were {\emph{not}} sampled from the Haar measure on the unitary group $U(2^n)$. Instead, they were randomly sampled (independently) as follows: For $j\in\{1,\dots , m+1\}$ we first chose a permutation $\tau_j\in S_n$ uniformly at random and subsequently (independently) sampled unitaries $U^{(j)}_1 ,\dots , U^{(j)}_{\left \lfloor{n/2}\right \rfloor }$ from the Haar measure on $SU(4)$. The unitary $C_j$ was then obtained by applying $U^{(j)}_1 ,\dots , U^{(j)}_{\left \lfloor{n/2}\right \rfloor }$ to the qubit pairs $(\tau_j (1),\tau_j (2)),\dots ,(\tau_j (2{\left \lfloor{n/2}\right \rfloor }-1),\tau_j (2{\left \lfloor{n/2}\right \rfloor }))$ respectively. Note that this is precisely how the individual layers in the quantum volume test \cite{quantum_volume} are sampled. For a visual representation of the circuits featuring in our experiments, see Figure \ref{fig:circuit}.

    \subsection{Alignment with Exact Gradient Vector}\label{subsec:alignment_exact_grad}
    Here we describe the experiments we carried out to compare the respective alignment of the noisy gradient and the denoised gradient with the exact gradient vector. In all experiments we proceeded as follows: We decided on a number of samples $N$ and fixed $n$, $m$, the number of shots per circuit evaluation, and the learning rate $\alpha = 0.1$. For each combination of $\ell$ and quantum backend featuring in the experiment, we then repeated the following $N$ (number of samples) times:
    \begin{enumerate}
        \item A circuit and a point $\theta_0\in\mathbb{R}^m$ are sampled randomly as explained in the beginning of Section \ref{sec:experiments},
        \item Algorithm \ref{algorithm:denoised_gradient_descent} is executed for $T=\ell$ steps (including gradient rescaling), and the last (denoised) gradient computed during the algorithm (at point $\theta_{\ell - 1}$) is denoted as $w_1\in\mathbb{R}^m$,
        \item The (noisy) gradient $w_2\in\mathbb{R}^m$ is computed at point $\theta_{\ell - 1}$ by way of the parameter-shift rules,
        \item Denoting the exact gradient at $\theta_{\ell - 1}$, computed via statevector simulation, as $w\in\mathbb{R}^m$, the cosine similarities $x_j = \frac{w^t w_j}{\Vert w\Vert \Vert w_j\Vert}$ between $w_j$ and $w$, where $j\in\{1,2\}$, are computed (the denominator is artificially bounded from below by a positive constant for the sake of numerical stability and in order to avoid division by zero),
        \item The point $(x_1, x_2)\in\mathbb [-1,1]\times [-1,1]$ is plotted in a coordinate system.
    \end{enumerate}
    We then end up with a scatter plot of $N$ points. The points below the diagonal $\{x=y\}$ correspond to outcomes where the denoised gradient obtained from Algorithm \ref{algorithm:denoised_gradient_descent} was closer to the exact gradient than the noisy gradient (where {\emph{closeness}} is measured in terms of cosine similarity; this is a sensible similarity measure, since the denoised gradient was rescaled to the length of the noisy gradient).

    Note that, for each of the $N$ samples, both circuit and point in parameter space are sampled randomly (independently), i.e., there is a large number of different circuits appearing in each experiment.

    We carry out two such experiments: In Section \ref{subsubsec:shot_noise} we focus on the effect of measurement shot noise, whereas in Section \ref{subsubsec:fake_devices} we focus on (simulated) quantum hardware noise.

    \subsubsection{Measurement Shot Noise}\label{subsubsec:shot_noise}
\begin{figure}[htp]
        \captionsetup{width=0.9\linewidth}
    	\centering
        \subcaptionbox{$\ell = 1$}{\includegraphics[width=2.23in]{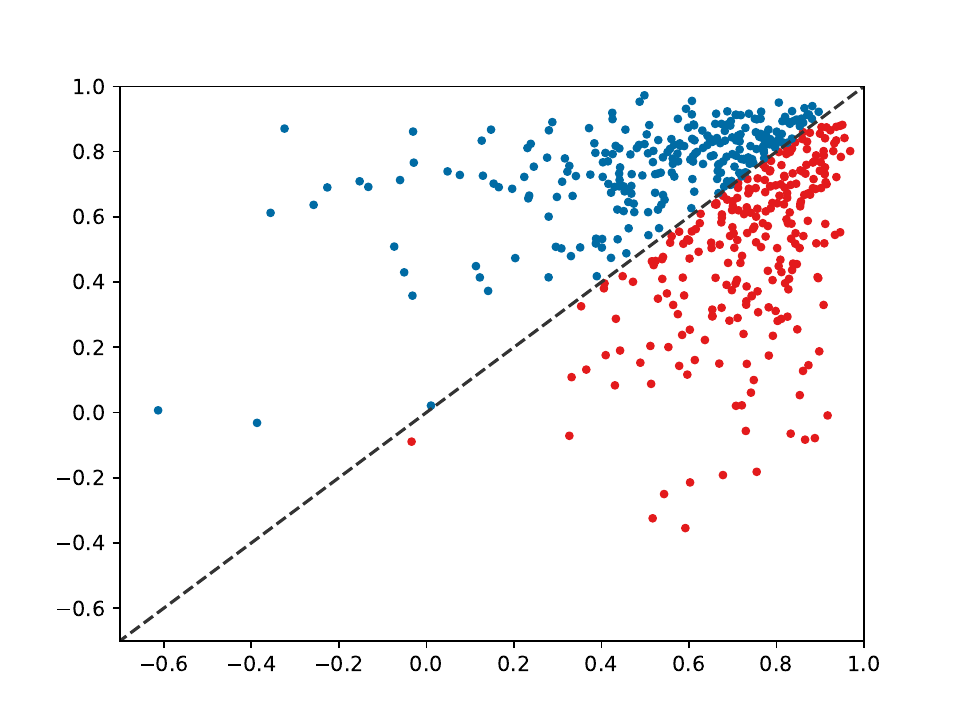}}\hspace{0.01em}%
        \subcaptionbox{$\ell = 2$}{\includegraphics[width=2.23in]{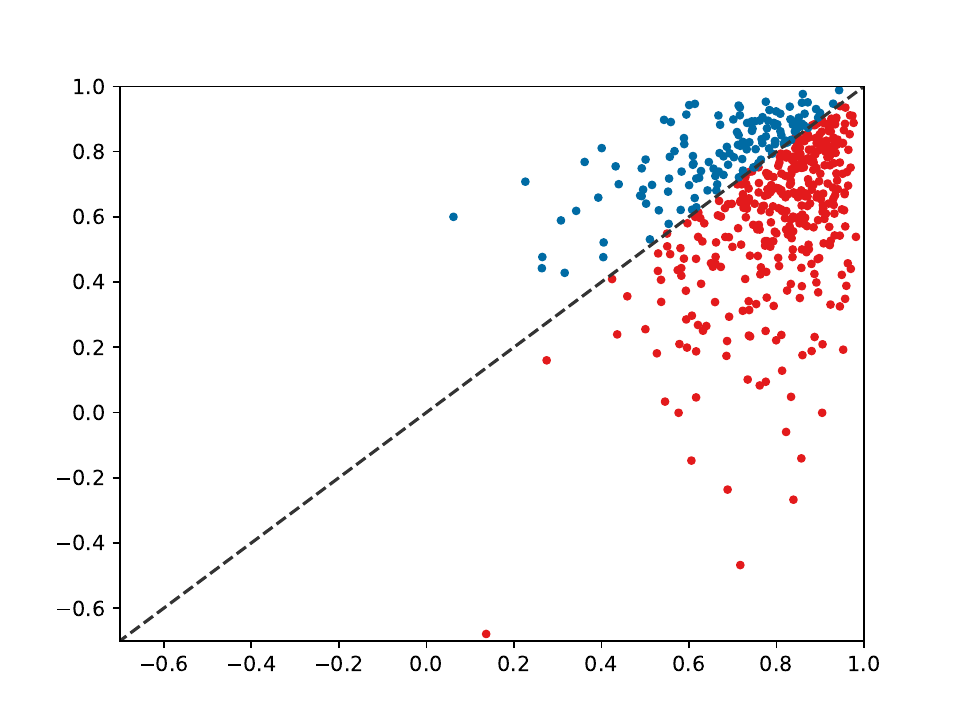}}\hspace{0.01em}%
        \subcaptionbox{$\ell = 3$}{\includegraphics[width=2.23in]{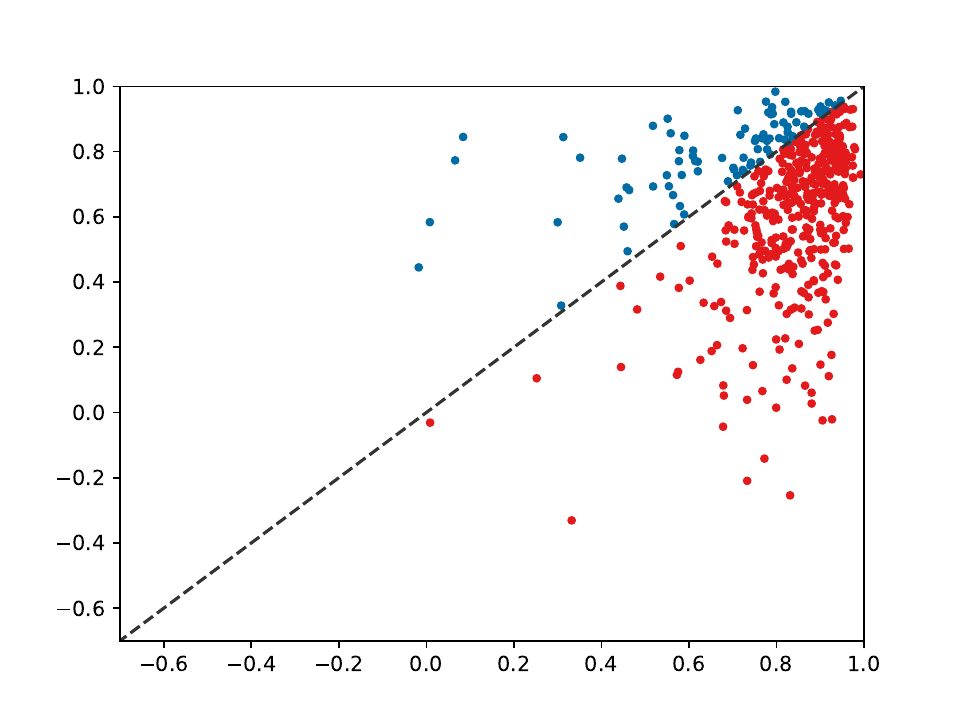}}
    	
    	\bigskip
    	
        \subcaptionbox{$\ell = 4$}{\includegraphics[width=2.23in]{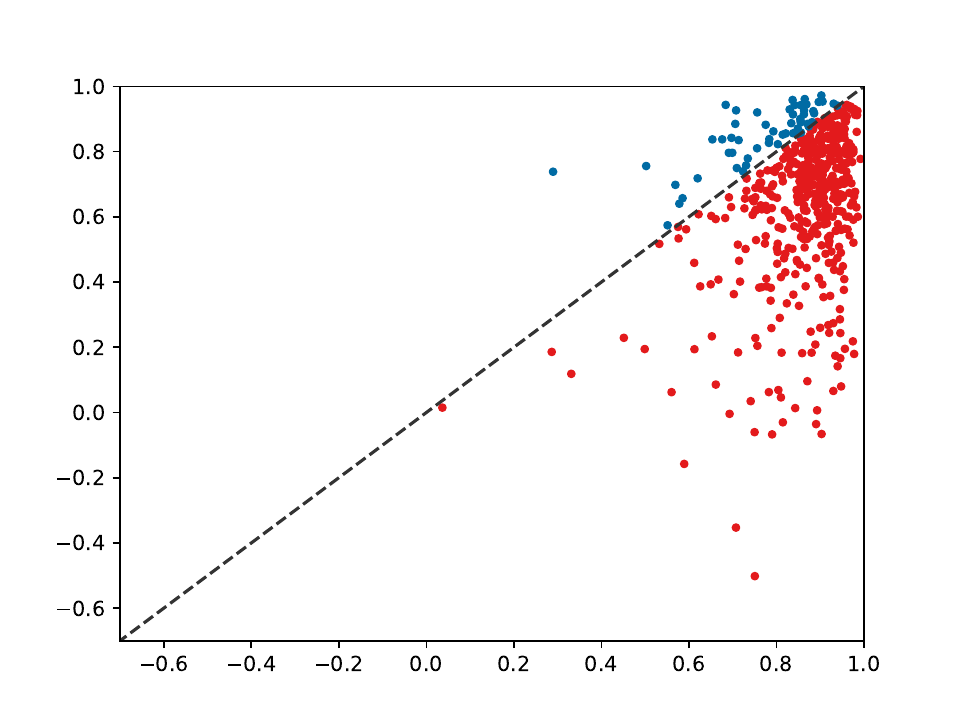}}\hspace{0.01em}%
        \subcaptionbox{$\ell = 5$}{\includegraphics[width=2.23in]{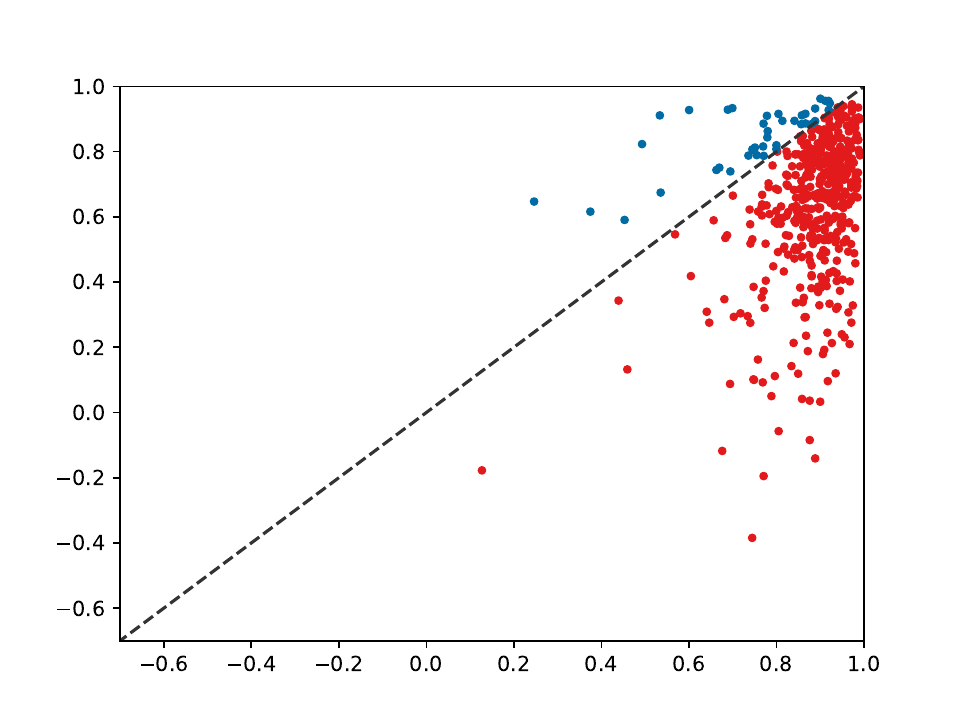}}\hspace{0.01em}%
        \subcaptionbox{$\ell = 6$}{\includegraphics[width=2.23in]{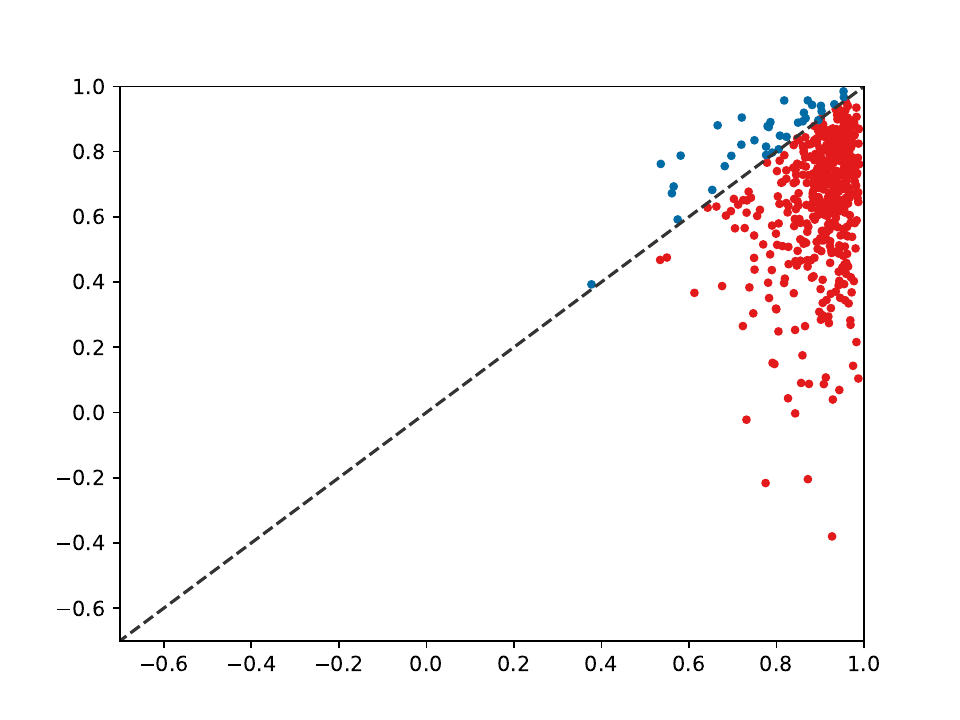}}
     
    	\caption{We investigated the effect that the choice of $\ell$ has on the ability of Algorithm \ref{algorithm:denoised_gradient_descent} to mitigate measurement shot noise. To that end, for each $\ell$, we randomly sampled $N=500$ circuits and points in parameter space and computed both a noisy and a denoised gradient (obtained from Algorithm \ref{algorithm:denoised_gradient_descent}). Subsequently, both of them were compared to the exact gradient (computed via statevector simulation): For each of the $N=500$ samples we plot a point in the $x$-$y$-plane, whose $x$- and $y$-coordinate are the cosine similarity of the exact gradient to the denoised gradient and to the noisy gradient respectively. Accordingly, points below the diagonal (red) correspond to outcomes where the denoised gradient obtained from Algorithm \ref{algorithm:denoised_gradient_descent} outperformed the noisy gradient; points on or above the diagonal (blue) correspond to outcomes where this was not the case. For more details, see Section \ref{subsubsec:shot_noise}.}
    	\label{fig:scatter_shot_noise}
    \end{figure}

    For this experiment we use $N=500$ samples and set $n=8$, $m=8$, $\lambda = 0.28$. Further, we set the number of measurement shots per circuit to $200$. Since we want to exclusively focus on measurement shot noise in this experiment, we choose the \verb|AerSimulator| (without noise model) provided by the Qiskit framework as the quantum backend for this experiment. We then carry out the procedure outlined in the beginning of Section \ref{subsec:alignment_exact_grad} with $\ell = 1,\dots ,6$, obtaining six scatter plots, see Figure \ref{fig:scatter_shot_noise}.
    
    For $\ell = 1,\dots ,6$, the denoised gradient obtained from Algorithm \ref{algorithm:denoised_gradient_descent} outperforms the noisy gradient in 50.0\%, 72.8\%, 82.6\%, 88.6\%, 91.8\%, 93.0\% of the cases respectively.

    \subsubsection{Quantum Hardware Noise}\label{subsubsec:fake_devices}
\begin{figure}[htp]
        \captionsetup{width=0.9\linewidth}
    	\centering
        \subcaptionbox{\texttt{FakeVigoV2}}{\includegraphics[width=2.23in]{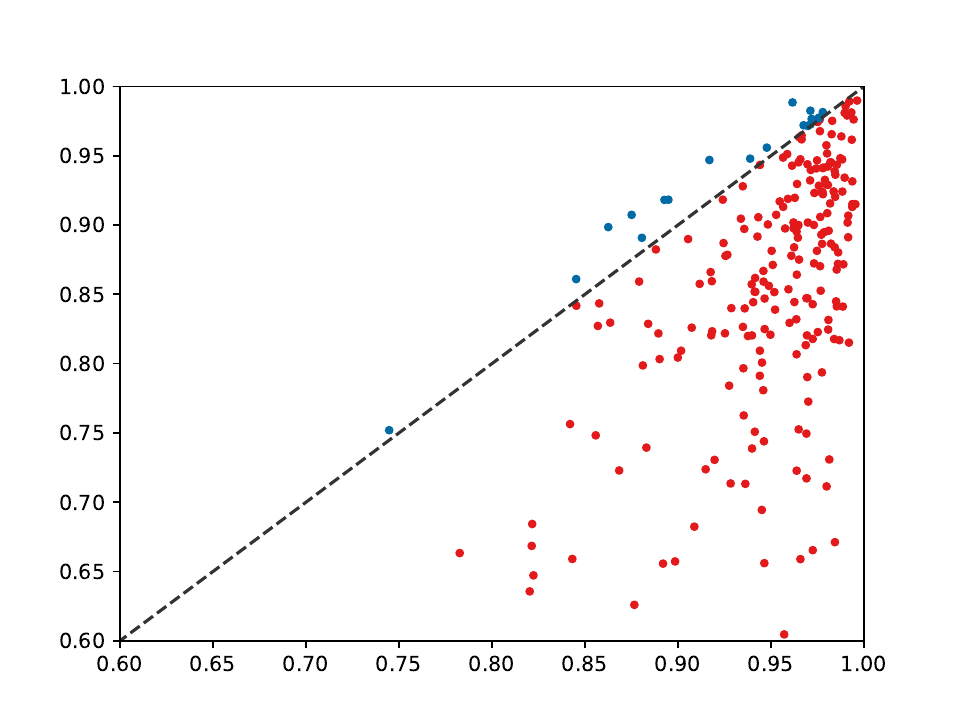}}\hspace{0.01em}%
        \subcaptionbox{\texttt{FakeNairobiV2}}{\includegraphics[width=2.23in]{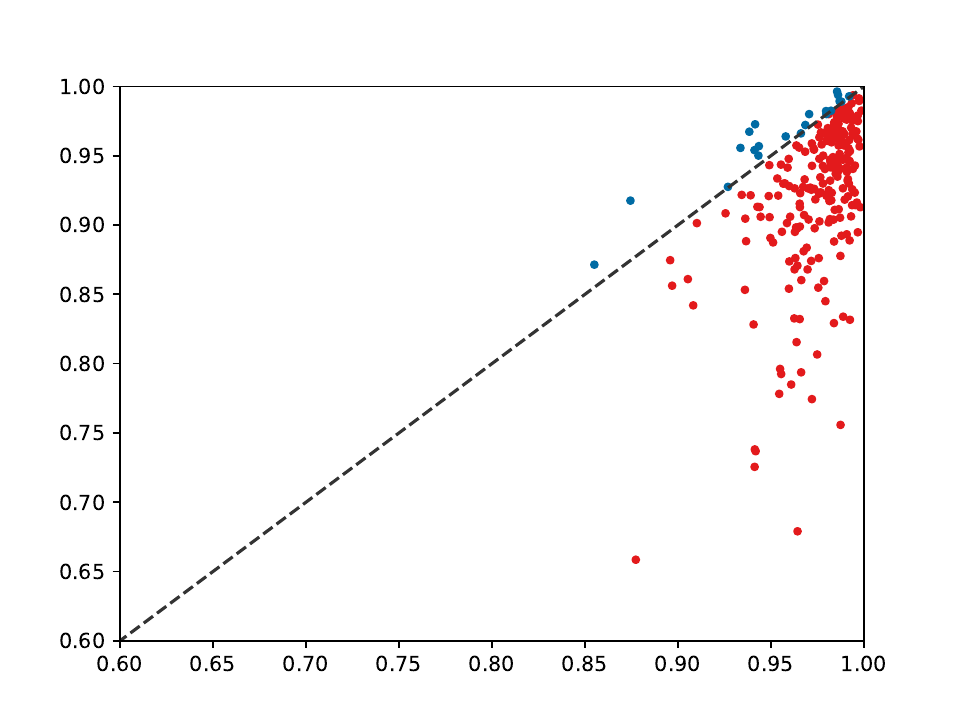}}\hspace{0.01em}%
        \subcaptionbox{\texttt{FakeCairoV2}}{\includegraphics[width=2.23in]{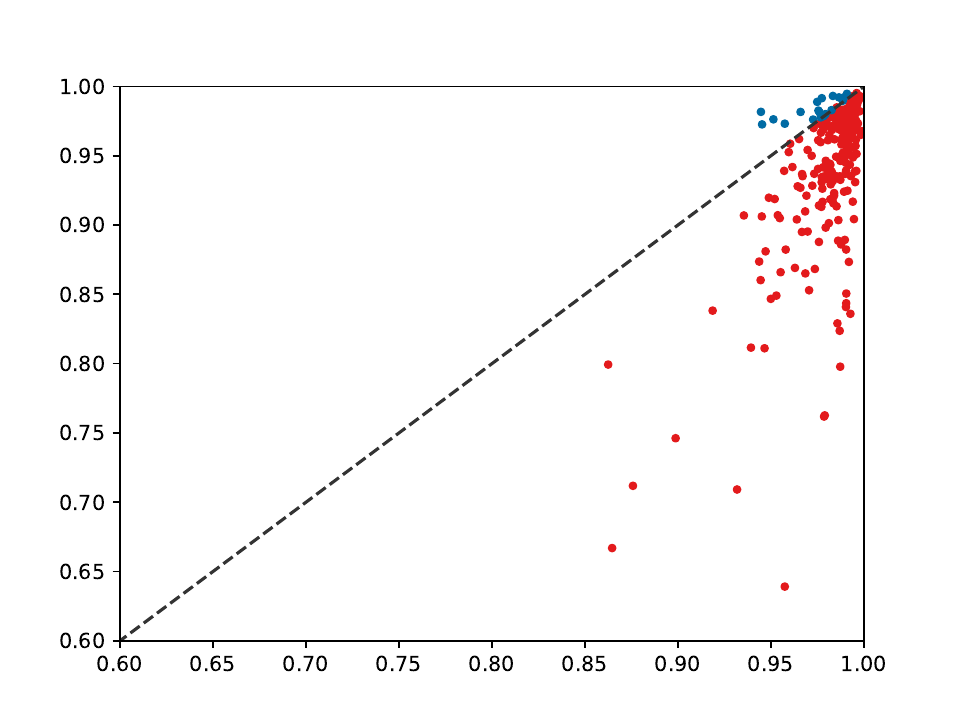}}
    	
    	\bigskip
    	
        \subcaptionbox{\texttt{FakeBrooklynV2}}{\includegraphics[width=2.23in]{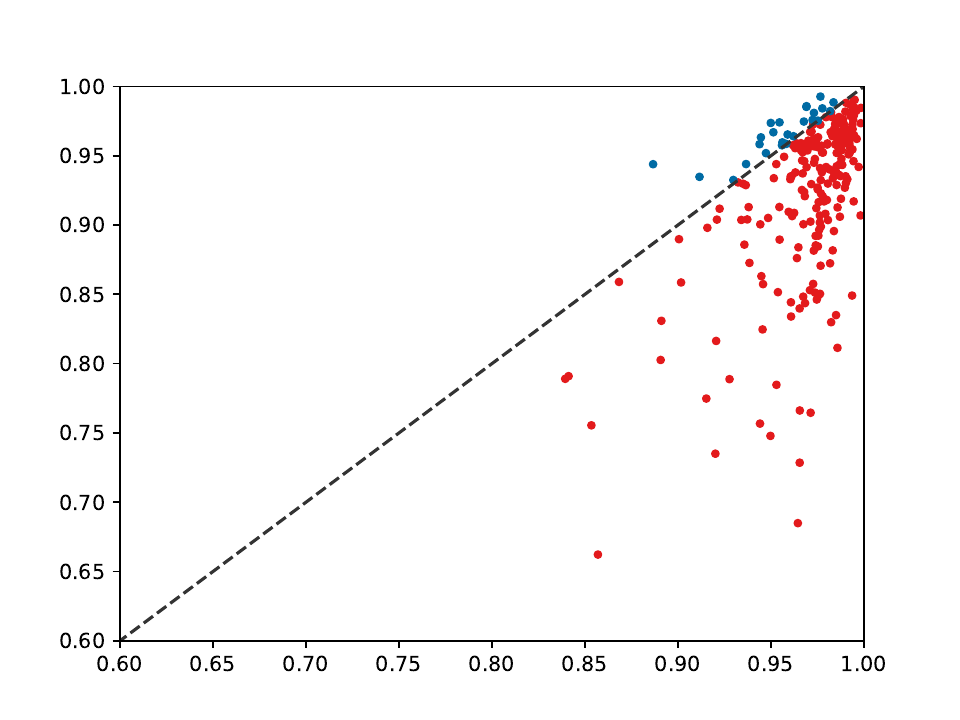}}\hspace{0.01em}%
        \subcaptionbox{\texttt{FakeWashingtonV2}}{\includegraphics[width=2.23in]{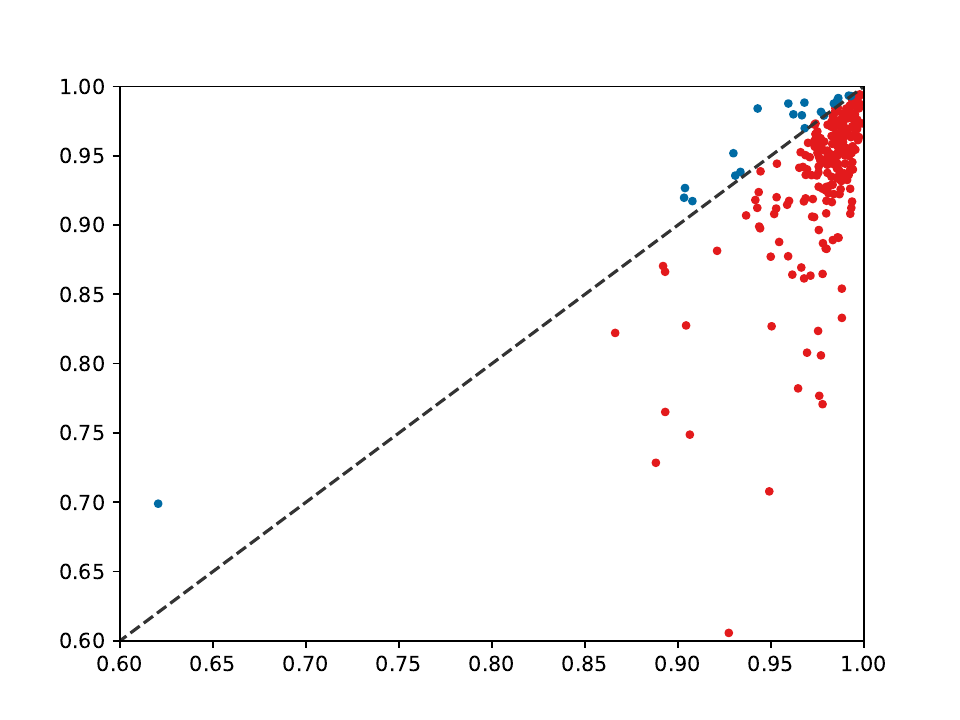}}\hspace{0.01em}%
        \subcaptionbox{\texttt{AerSimulator} (without noise model)}{\includegraphics[width=2.23in]{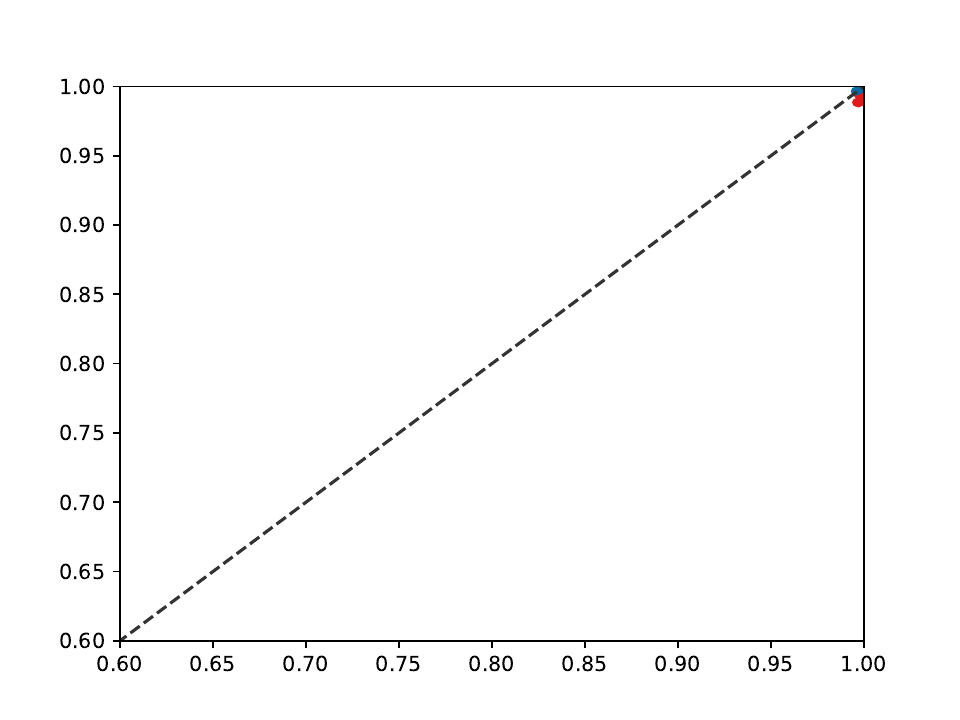}}
     
    	\caption{We investigated the ability of Algorithm \ref{algorithm:denoised_gradient_descent} to mitigate the effect of quantum hardware noise. To that end, for each quantum backend, we randomly sampled $N=250$ circuits and points in parameter space and computed both a noisy and a denoised gradient (obtained from Algorithm \ref{algorithm:denoised_gradient_descent}). Subsequently, both of them were compared to the exact gradient (computed via statevector simulation): For each of the $N=250$ samples we plot a point in the $x$-$y$-plane, whose $x$- and $y$-coordinate are the cosine similarity of the exact gradient to the denoised gradient and to the noisy gradient respectively. Accordingly, points below the diagonal (red) correspond to outcomes where the denoised gradient obtained from Algorithm \ref{algorithm:denoised_gradient_descent} outperformed the noisy gradient; points on or above the diagonal (blue) correspond to outcomes where this was not the case. The \texttt{AerSimulator} backend (without noise model) was included in order to demonstrate that the effect of measurement shot noise is negligible in this experiment. For clarity of visual presentation, we only show points for which both $x$- and $y$-coordinate are $\geq 0.6$ (all points which are not shown are red, i.e., correspond to outcomes where the denoised gradient outperformed the noisy gradient). For more details, see Section \ref{subsubsec:fake_devices}.}
    	\label{fig:scatter_fake_devices}
    \end{figure}

    For this experiment we use $N=250$ samples and set $n=5$, $m=8$, $\lambda = 0.04$, and $\ell = 5$. In order to suppress the effect of measurement shot noise, we set the number of measurement shots per circuit to $10000$. We then carry out the procedure outlined in the beginning of Section \ref{subsec:alignment_exact_grad} with the (simulated) quantum hardware backends \verb|FakeVigoV2|, \verb|FakeNairobiV2|, \verb|FakeCairoV2|, \verb|FakeBrooklynV2|, \verb|FakeWashingtonV2| provided by the Qiskit framework. In order to demonstrate that measurement shot noise plays a negligible role in this experiment, we also carry out the above-mentioned procedure with the \verb|AerSimulator| (without noise model) provided by the Qiskit framework. We thus obtain a total of six scatter plots, see Figure \ref{fig:scatter_fake_devices}.

    As expected, for the \verb|AerSimulator| (without noise model), all noisy and denoised gradients were very close to the exact gradient. For the (simulated) quantum hardware backends \verb|FakeVigoV2|, \verb|FakeNairobiV2|, \verb|FakeCairoV2|, \verb|FakeBrooklynV2|, \verb|FakeWashingtonV2|, the denoised gradient obtained from Algorithm \ref{algorithm:denoised_gradient_descent} outperforms the noisy gradient in 92.4\%, 91.6\%, 92.4\%, 89.6\%, 91.6\% of the cases respectively.

    \subsection{Descent of Objective Function}\label{subsec:descent_objective}
    In order to keep the computational load manageable, we set $n=4$, $m=4$, $\ell = 5$, learning rate $\alpha = 0.4$, number of steps $T=60$. We further set the number of measurement shots per circuit to $50$. We then randomly sample a single circuit and initial point $\theta_0\in\mathbb{R}^m$ in parameter space as described in the beginning of Section \ref{sec:experiments}. For each combination of regularization hyperparameter $\lambda$ and quantum backend featuring in this experiment, we then repeat the following $N=100$ times:
    \begin{enumerate}
        \item Execute Algorithm \ref{algorithm:denoised_gradient_descent} with initial point $\theta_0$ (including gradient rescaling),
        \item Execute noisy gradient descent with initial point $\theta_0$ using the parameter-shift rules with the same number of steps ($T= 60$), the same learning rate ($\alpha = 0.4$), and the same number of measurement shots per circuit ($50$).
    \end{enumerate}
    For both algorithms we thus obtain $N=100$ sequences of points $(\theta_0 , \dots , \theta_{60})$ in parameter space. For the sake of better comparability we then use statevector simulation to evaluate the {\emph{exact}} values of $f$ (see Section \ref{subsec:algorithm_setting}) at these points, which, for both algorithms, yields $N=100$ sequences of values $(f(\theta_0 ), \dots , f(\theta_{60}))$. It is important to point out that exact evaluations of the function $f$ were not employed {\emph{during}} the execution of either algorithm, but rather {\emph{after}} the executions of the algorithms in order to compare the results. I.e., exact evaluations of $f$ did not have an impact on the execution of either algorithm.

    For both algorithms we then computed the component-wise average of the $N=100$ sequences of values, giving precisely one sequence of values of length $T+1=61$ for each algorithm. These two sequences can be interpreted as the respective average performance of Algorithm \ref{algorithm:denoised_gradient_descent} and noisy gradient descent. For comparison sake we then computed a corresponding sequence of $61$ points using exact gradient descent (same initial point, number of steps, and learning rate as for noisy gradient descent, but the gradients were computed via the parameter-shift rules using statevector simulation).

\begin{figure}[htp]
        \captionsetup{width=0.9\linewidth}
    	\centering
        \subcaptionbox{\texttt{FakeVigoV2}, Algorithm \ref{algorithm:denoised_gradient_descent} with $\lambda = 0.01/\sqrt{50}$ compared to noisy and exact gradient descent}{\includegraphics[width=2.138in]{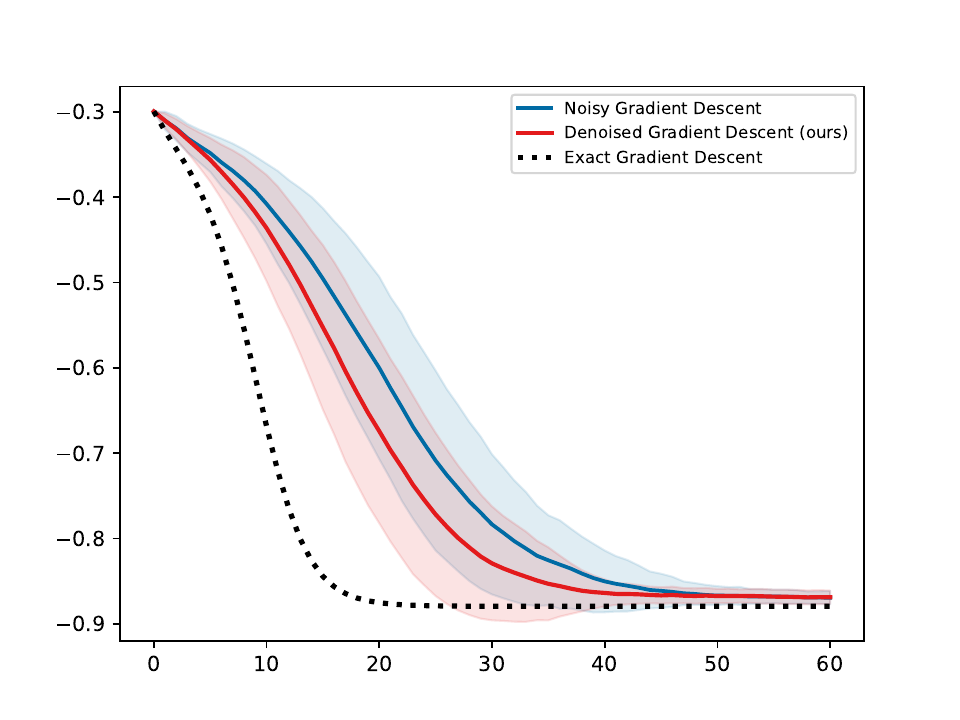}}\hspace{1.0em}%
        \subcaptionbox{\texttt{FakeNairobiV2}, Algorithm \ref{algorithm:denoised_gradient_descent} with $\lambda = 0.01/\sqrt{50}$ compared to noisy and exact gradient descent}{\includegraphics[width=2.138in]{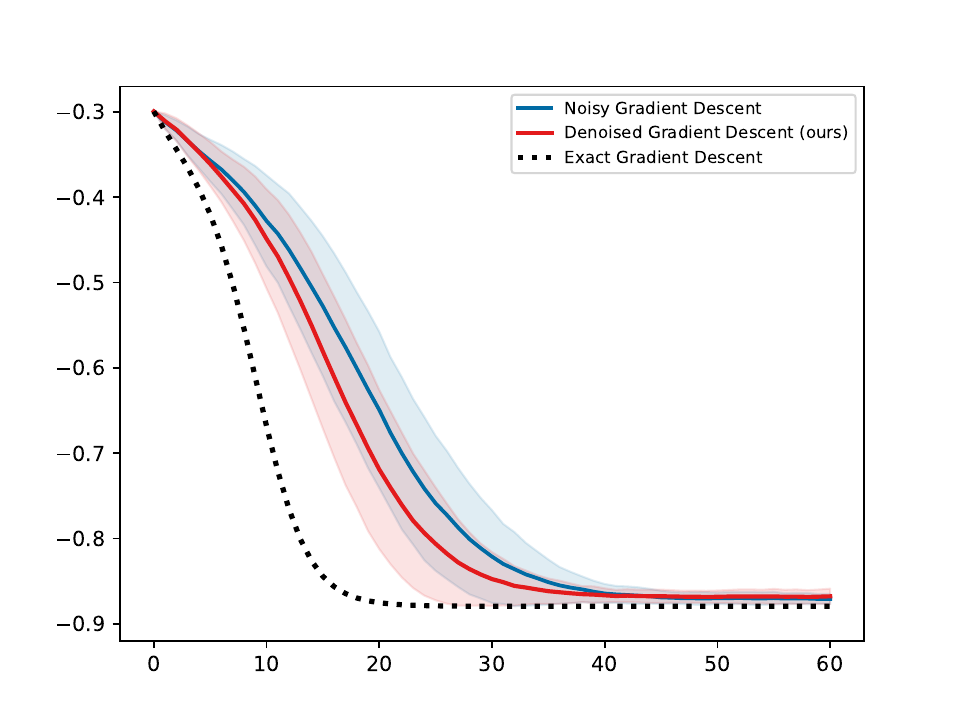}}\hspace{1.0em}%
        \subcaptionbox{\texttt{FakeCairoV2}, Algorithm \ref{algorithm:denoised_gradient_descent} with $\lambda = 0.01/\sqrt{50}$ compared to noisy and exact gradient descent}{\includegraphics[width=2.138in]{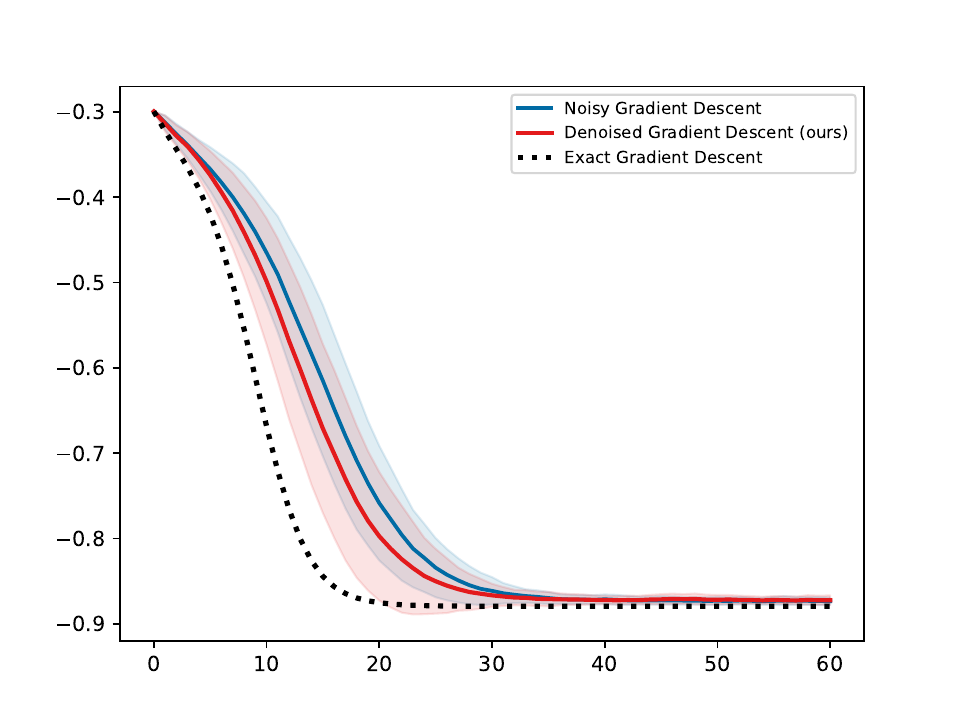}}
    	
    	\bigskip
    	
        \subcaptionbox{\texttt{FakeVigoV2}, Algorithm \ref{algorithm:denoised_gradient_descent} with different values for $\lambda$}{\includegraphics[width=2.138in]{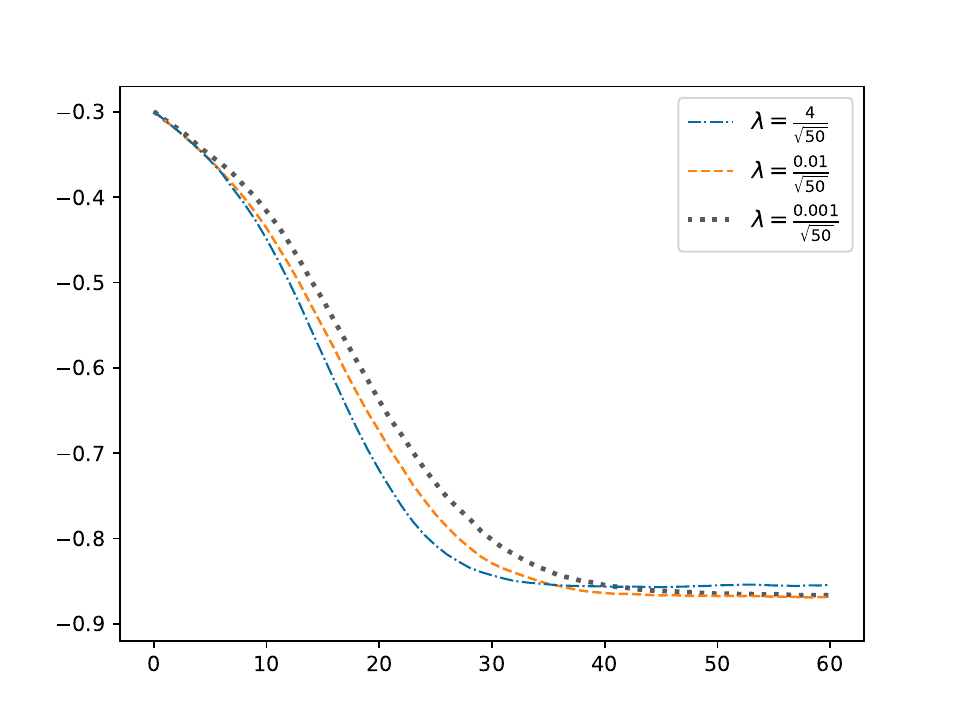}}\hspace{1.0em}%
        \subcaptionbox{\texttt{FakeNairobiV2}, Algorithm \ref{algorithm:denoised_gradient_descent} with different values for $\lambda$}{\includegraphics[width=2.138in]{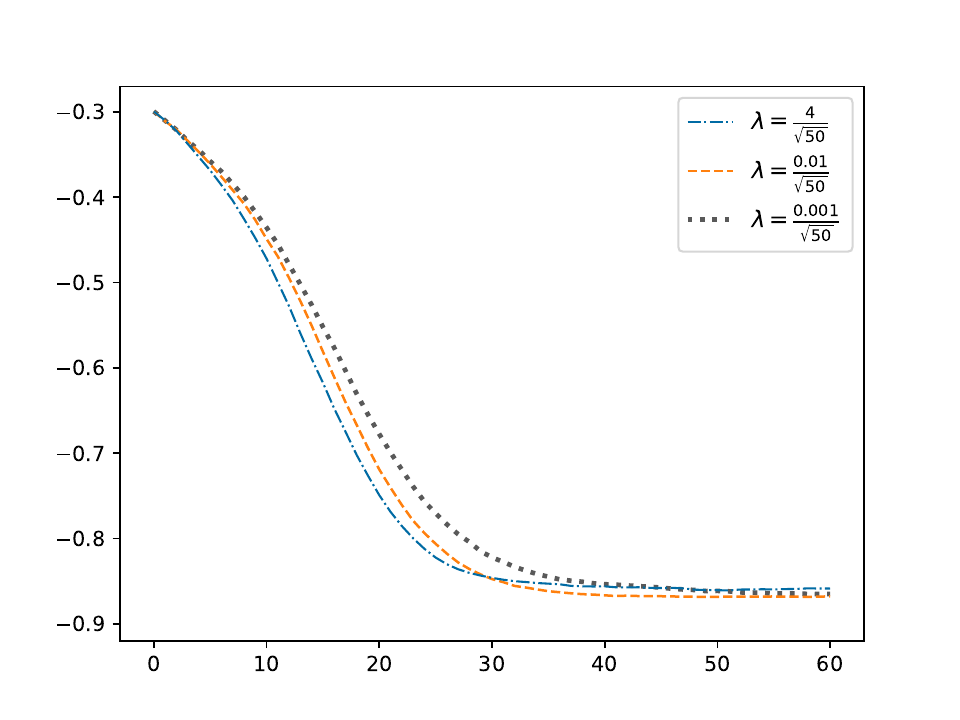}}\hspace{1.0em}%
        \subcaptionbox{\texttt{FakeCairoV2}, Algorithm \ref{algorithm:denoised_gradient_descent} with different values for $\lambda$}{\includegraphics[width=2.138in]{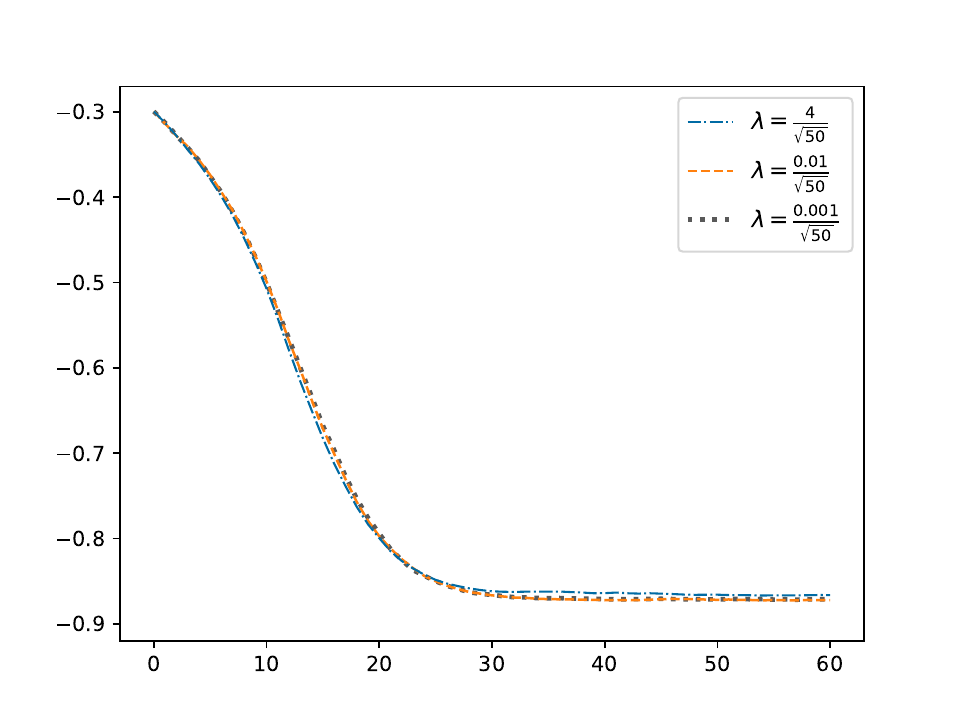}}
     
    	\caption{We executed Algorithm \ref{algorithm:denoised_gradient_descent} on several (simulated) quantum hardware backends and using several values for the regularization hyperparameter $\lambda$. The circuit and the initial point in parameter space were sampled randomly. For the sake of comparison, we also executed exact gradient descent (using statevector simulation) and noisy gradient descent using the parameter-shift rules. Algorithm \ref{algorithm:denoised_gradient_descent} was executed $N=100$ times for each combination of $\lambda$ and quantum hardware backend. Noisy gradient descent was executed $N=100$ times for each quantum hardware backend. The resulting families of $N=100$ curves were averaged respectively; the standard deviation is indicated in some of the plots. For more details, see Section \ref{subsec:descent_objective}.}
    	\label{fig:lossfct}
    \end{figure}

    This procedure was carried out for each of the $9$ combinations of regularization hyperparameters $\lambda = 4/\sqrt{50}$, $\lambda = 0.01/\sqrt{50}$, and $\lambda =0.001/\sqrt{50}$ with (simulated) quantum hardware backends \verb|FakeVigoV2|, \verb|FakeNairobiV2|, \verb|FakeCairoV2|. The results are visualized in Figure \ref{fig:lossfct}.

    The results shown in Figure \ref{fig:lossfct} validate our algorithm. However, as expected, there is some indication that the optimal choice for the regularization hyperparameter $\lambda$ might be device-dependent. Non-surprisingly, the results seem to further indicate that -- for optimal performance -- $\lambda$ should be adjusted over the course of the algorithm (e.g., based on the length of the noisy gradient vector).

    \section{Discussion}\label{sec:discussion}

    In this section we will discuss both the advantages and the drawbacks of our algorithm compared to (noisy) gradient descent.

    The obvious advantage is that, in many scenarios in the context of variational quantum algorithms, our denoised gradient descent algorithm is able to significantly accelerate the descent of the objective function and to improve the alignment of the estimated gradient vector with the exact gradient vector when compared to noisy gradient descent. Several scenarios where this is indeed the case were explored in Section \ref{sec:experiments}. Moreover, the computational overhead of our algorithm is entirely classical -- the number of circuit evaluations is exactly the same as when executing (noisy) gradient descent using the parameter-shift rules.
    
    However, the denoised gradient descent algorithm comes with some caveats, the most obvious one being that, since our algorithm makes use of samples from past iterations, it is not really suitable for variational quantum algorithms which take data as input, i.e., for those variational quantum algorithms whose corresponding ansatz has parametrized gates corresponding to both inputs and trainable parameters (our algorithm would still prove beneficial if one was to carry out several consecutive gradient descent steps with the same mini batch of training data, but this is a niche application that we will not consider here). Instead, our algorithm is well-suited for variational quantum algorithms whose ansatz only contains parametrized gates corresponding to trainable parameters; the most prominent examples of the latter are variational quantum eigensolvers \cite{VQE}.
    
    There are also caveats regarding the performance resp.\ feasibility of the algorithm. For example, if the number of trainable parameters $m$ or the number of iterations $\ell$ to consider when computing the approximation becomes too large, it might become infeasible to solve the linear system of equations appearing in Algorithm \ref{algorithm:denoised_gradient_descent}: In each iteration $t\geq\ell$, the latter will be a square linear system of equations with $2m\ell$ unknowns resp.\ equations. 
    Furthermore, while we expect it to be straightforward to establish a rigorous advantage in the case of measurement shot noise (under reasonable assumptions), one cannot expect our algorithm to offer a tangible advantage for all kinds of hardware quantum noise -- a detailed analysis is left for future work. In this context it is also important to mention that the optimal choice of the hyperparameter $\lambda >0$ does not only depend on the number of measurement shots per circuit, but also on the quantum device on which the circuits are executed. As such, good heuristics for the choice of $\lambda$ will necessarily be device-dependent. Finding good heuristics for the choice of $\lambda$ is further complicated by the fact that -- for optimal performance -- $\lambda$ should be adjusted over the course of the algorithm (e.g., based on the length of the noisy gradient vector).
    
    Finally, we mention that the advantage offered by our algorithm might disappear if the learning rate is chosen too large. This is because the quality of the (local) approximation computed by our algorithm declines if the points at which the objective function $f$ is sampled are spaced further apart.

    When weighing the advantages and drawbacks outlined above, we believe that there are some scenarios with practical relevance where using our algorithm would prove advantageous.

    \section{Conclusion and Outlook}\label{sec:conclusion}
    In this article we introduced the denoised gradient descent algorithm, which mitigates the effect of noise on gradient descent in variational quantum algorithms. We explored the capabilities of the algorithm experimentally and discussed its advantages and drawbacks. Potential topics for future work include, but are not limited to, the following:
    \begin{itemize}
        \item deriving rigorous performance guarantees for the algorithm under suitable assumptions on the noise,
        \item thoroughly analyzing our algorithm for different types of noise channels,
        \item deriving good device-dependent heuristics for the choice of the hyperparameter $\lambda >0$ (including heuristics for adjusting $\lambda$ over the course of the algorithm),
        \item determining how large the learning rate can be chosen without the quality of the local approximation deteriorating to the point where using our algorithm is no longer advantageous,
        \item investigating whether our algorithm can be used to reduce the amount of measurement shots necessary to succesfully carry out gradient descent in variational quantum algorithms, see for example \cite{scriva2023challenges}. The experimental results in Section \ref{sec:experiments} seem to indicate that this is possible -- however, further studies are needed to verify this.
    \end{itemize}

    \bibliographystyle{alpha}
	\bibliography{literature/bibliography}

\end{document}